\documentclass[aps,prb,twocolumn,showpacs,preprintnumbers,amsmath,amssymb,superscriptaddress]{revtex4}
\usepackage{graphicx}
\usepackage{dcolumn}
\usepackage{bm}

\newcommand{\ket}[1]{\ensuremath{\left|{#1}\right\rangle}}
\newcommand{\bra}[1]{\ensuremath{\left\langle{#1}\right|}}
\newcommand{\braket}[1]{\ensuremath{\left\langle{#1}\right\rangle}}

\begin{document}

\title{Dynamics of entanglement via propagating microwave photons}

\author{C. Sab\'in}
\affiliation{Instituto de F\'{\i}sica Fundamental, CSIC,
  Serrano 113-B, 28006 Madrid, Spain}
\email{csl@imaff.cfmac.csic.es}

\author{J.~J. Garc{\'\i}a-Ripoll}
\affiliation{Instituto de F\'{\i}sica Fundamental, CSIC,
  Serrano 113-B, 28006 Madrid, Spain}

\author{E. Solano}
\affiliation{Departamento de Qu\'{\i}mica F\'{\i}sica, Universidad del Pa\'{\i}s Vasco - Euskal Herriko Unibertsitatea, Apdo.\ 644, 48080 Bilbao, Spain}
\affiliation{IKERBASQUE, Basque Foundation for Science, Alameda Urquijo 36, 48011 Bilbao, Spain}

\author{J. Le\'on}
\affiliation{Instituto de F\'{\i}sica Fundamental, CSIC,
  Serrano 113-B, 28006 Madrid, Spain}

\date{\today}

\begin{abstract}
We propose a simple circuit quantum electrodynamics (QED) experiment to test the generation of entanglement between two superconducting qubits. Instead of the usual cavity QED picture, we study qubits which are coupled to an open transmission line and get entangled by the exchange of propagating photons. We compute their dynamics using a full quantum field theory beyond the rotating-wave approximation and explore a variety of regimes which go from a weak coupling to the recently introduced ultrastrong coupling regime. Due to the existence of single photons traveling along the line with finite speed, our theory shows a light cone dividing the spacetime in two different regions. In one region, entanglement may only arise due to correlated vacuum fluctuations, while in the other the contribution from exchanged photons shows up.
\end{abstract}


\maketitle
\section{Introduction}

Quantum mechanics does not allow us in general to consider two arbitrary distant systems as separate~\cite{schrodinger}. In some cases there exist quantum correlations that cannot be generated by local operations and classical communication between remote systems. Time enters this picture through two different questions. The first one is related to the speed bound of a hypothetical superluminal influence which could explain all quantum correlations, estimated to be $10^4 c$ in a recent experiment~\cite{salart08}. The second question is of a more practical nature inside the quantum theory~\cite{reznik,reznikII,franson,conjuan,conjuanII}: what is the speed at which two distant systems become entangled?

Quantum field theory (QFT) fulfills the principle of microscopic causality by which two space-like separated events cannot influence each other~\cite{microcausalidad} and thus cannot be used to transfer information~\cite{power97,gisin}. We may then ask whether microcausality also sets a limit on the speed at which entanglement can be created between two separate systems. More precisely, can two subsystems, supported at regions $(\mathbf{x},t)$ and $(\mathbf{x}',t')$, become entangled while they are still space-like separated? Or in simple terms, can finite quantum correlations develop before signals arrive? The answer to this far reaching question is yes, it is possible. After all, Feynman propagators are finite beyond the light cone and even before photon arrival there exist correlations between the vacuum fluctuations at any two space-like separated events.

In this work we demonstrate that circuit QED is arguably one of the most suitable fields to study the dynamics of entanglement between distant systems. One reason is the existence of various choices of high quality superconducting qubits, the so-called artificial atoms~\cite{martinis85,bouchiat98,mooij99,transmon}. Another reason is the possibility of coupling those qubits strongly with traveling photons using microwave guides and cavities~\cite{blais04,wallraff04,chiorescu04}. Furthermore, those coupling strengths can reach the ultrastrong coupling regime~\cite{abdumalikov08,bourassa09,guenter09}, where the qubit-photon interaction approaches the energies of the qubit and photons. In this case, the rotating-wave approximation (RWA) breaks down and a different physical structure emerges. Such regimes can be activated and deactivated~\cite{peropadre09}, facilitating the creation of a fairly large amount of entanglement in a time-dependent way, as we will see in this work.

\begin{figure}[b]
  \centering
  \includegraphics[width=0.95\linewidth]{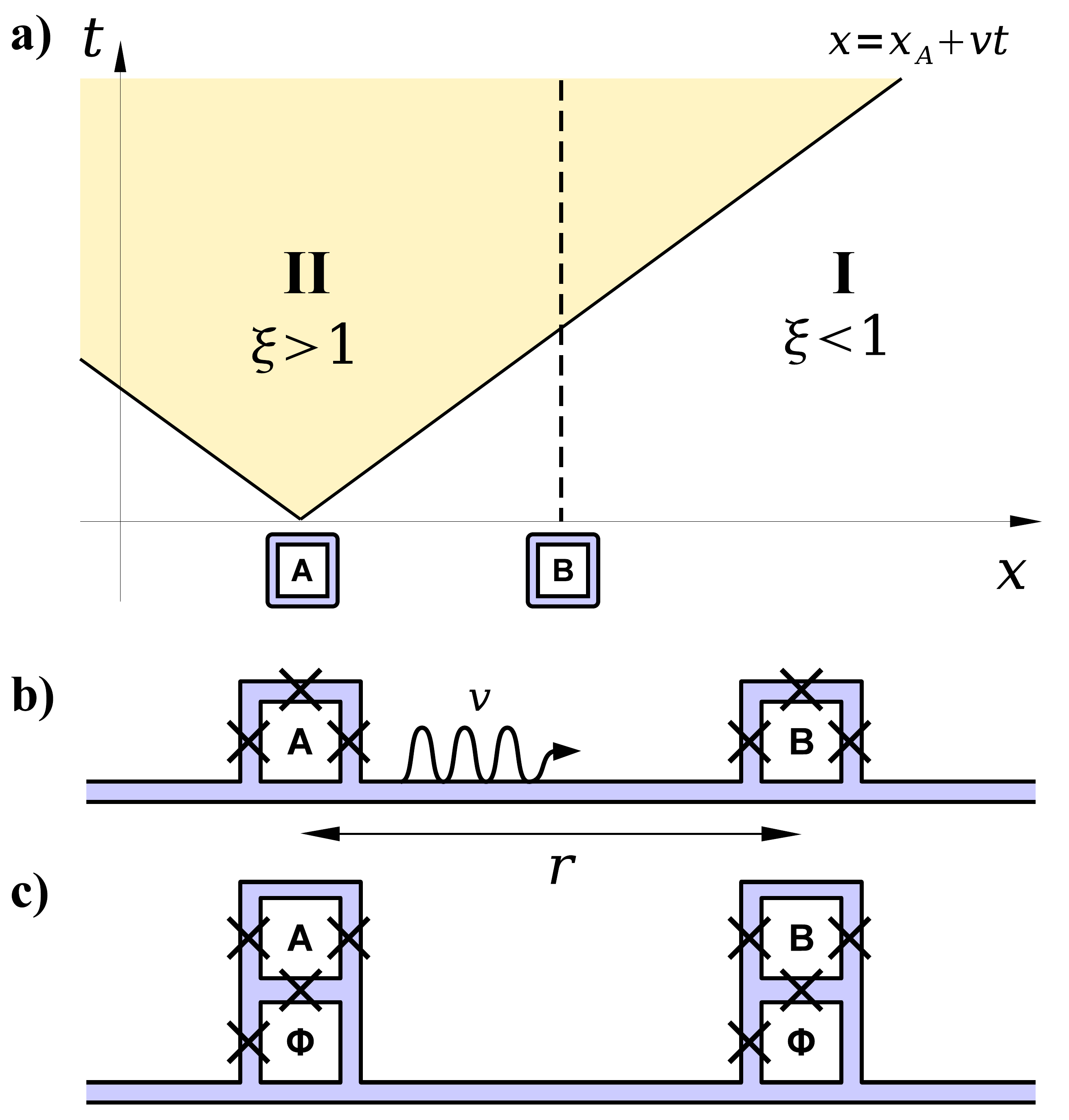}
  \caption{(a) Qubits that interact via traveling photons with finite velocity $v$ can be space-like (I, white) or time-like (II, shaded) separated, depending on the value of $\xi=vt/r.$ While only in II they are causally connected, entanglement may appear already in region I. (b) A possible implementation of these ideas consists of flux qubits ultrastrongly coupled to a common transmission line. (c) With a slight modification, the coupling of the qubits to the line can be dynamically tuned via fast magnetic fluxes, $\Phi$ (Color online). }
  \label{fig:setup}
\end{figure}

We will discuss some of the preceding questions in the framework of a precise circuit QED setup, see Fig.~\ref{fig:setup}, consisting on two well separated superconducting qubits coupled ultrastrongly to an open transmission line [Fig.~\ref{fig:setup}b]. The waveguide provides a continuum of microwave photons propagating with uniform velocity, $v,$ mediating an interaction between the qubits. Given an intitial separable state in which only qubit A is excited, we have studied the evolution of correlations and related it to the propagation of photons between qubits. The main results are: (i) Outside the light cone, that is in region I of Fig.~\ref{fig:setup}a where $\xi=vt/r < 1$, the excitation probability of qubit B is independent of the distance $r$ to qubit A. (ii) Still in region I, entanglement between the qubits always takes a finite value and grows with time. (iii) Once the qubits are time-like separated, that is as soon as we cross into region II, entanglement grows faster than the excitation probability of qubit B and takes sizeable values. Result (i) is a manifestation of the fact that our Quantum Field Theory (QFT) model satisfies microscopic causality, which formally translates into the vanishing of commutators associated with observables at space-like separations, $[{\mathcal{Q}}(\mathbf{x},t),{\mathcal{Q'}}(\mathbf{x}',t')]= 0 $ for $|\mathbf{x}-\mathbf{x}'|^2-c^2(t-t')^2 >0$. Furthermore, it shows that two qubits which are space-like separated cannot be used to communicate superluminal information. Result (ii), on the other hand, reveals the fact that correlations between vacuum fluctuations at separate points can be established at arbitrarily short times, even though they are non-signalling and cannot transmit information.

It is important to remark that the previous questions have been posed theoretically using model detectors \cite{reznik}, two-level atoms \cite{franson,conjuan,conjuanII}, scalar fields \cite{reznik,reznikII} and photons \cite{franson,conjuan,conjuanII}, yet no experimental test has been accomplished. However in this work, we show that the access to the ultrastrong couplings in circuit QED allows us to explore these ideas with very advantageous parameter ranges.

\section{Superconducting qubits coupled to a quantum field}

Our setup consists of two qubits, $A$ and $B,$ interacting via a quantum electromagnetic field. The qubits have two stationary states $\ket{e}$ and $\ket{g}$ separated by an energy $\hbar\Omega$ and interact with a one-dimensional field, which propagates along the line connecting them,
\begin{eqnarray}
  V(x)=\int dk \sqrt{N\omega_k}\left[e^{ikx}a_k +\mathrm{H.c.}\right].\label{field}
\label{field}
\end{eqnarray}
This field is described by a continuum of Fock operators $[a_k,a^{\dag}_{k'}]=\delta_{kk'},$ and a linear spectrum, $\omega_k = v|k|,$ where $v$ is the propagation velocity of the field and plays the role of the speed of light. The normalization and the speed of
photons depend on the microscopic details. In particular $v=1/\sqrt{cl},$ where $c$ and $l$ are the capacitance and inductance per unit length.

We consider qubits that are much smaller than the relevant wavelengths, $\lambda=v/\Omega,$ and lay well separated. Under these conditions we can split the Hamiltonian, $H = H_0 + H_I,$ into a free part for the qubits and the field
\begin{equation}
  H_0 = \frac{1}{2}\hbar\Omega(\sigma^z_A + \sigma^z_B) + \sum_k \omega(k)
  a^{\dagger}_ka_k,
\end{equation}
and a point-like interaction between them
\begin{equation}
  H_I = \sum_{\alpha=A,B} d_\alpha\,V(x_\alpha). \label{interaction}
\end{equation}
Here $x_A$ and $x_B$ are the fixed positions of the atoms, and $d_n=d\times \sigma^x_n$ is equivalent to the dipole moment in the case of atoms interacting with the electromagnetic field.

In what follows we choose the initial state $|\psi(0)\rangle = |eg\rangle\otimes|0\rangle,$ where only qubit $A$ has been excited, while both $B$ and the field remain in their ground and vacuum states, respectively. In the interaction picture given by the ``free'' Hamiltonian $H_0,$ the system evolves during a lapse of time $t$ into the state
\begin{equation}
  \ket{\psi(t)} = {\cal T}[e^{-i \int_0^tdt' H_I(t')/\hbar}]\ket{eg}\otimes\ket{0},\label{c}
\end{equation}
${\cal T}$ being the time ordering operator. Up to second order in perturbation theory the final state can be written as
\begin{eqnarray}
  \ket{\psi(t)}  = && \!\!\!\!\!  \left[(1+A)\ket{eg} + X\ket{ge}\right]\otimes\ket{0} +
  \label{wavefunction} \nonumber \\
 &&  (U_A\ket{gg} + V_B\ket{ee})\otimes\ket{1} \nonumber \\ && + (F\ket{eg} +  G \ket{ge})\otimes\ket{2} + {\cal O}(d^3).
\end{eqnarray}
The coefficients for the vacuum, single-photon, and two-photon states,
are computed using the action $(\alpha=A,B)$
\begin{eqnarray}
  \mathcal{S}^+_\alpha \! = \! - \frac{i}{\hbar}
  \int_0^t
  e^{i\Omega t'}\braket{e_\alpha|d\sigma^x_\alpha|g_\alpha} V(x_\alpha,t') dt'
  = -(\mathcal{S}^{-}_\alpha)^\dagger\label{f}
\end{eqnarray}
among different photon number states $\ket{n}, n=0,1,2\ldots$, being $\ket{n}\bra{n}=\frac{1}{n!}\int dk_1....\int dk_n\ket{k_1...k_n}\bra{k_1...k_n}$ and $\ket{k}=a_k^{\dagger}\ket{0}$. Only one term corresponds to interaction
\begin{equation}
  X = \langle0|T(\mathcal{S}^+_B \mathcal{S}^-_A)|0\rangle . \label{exchange}
\end{equation}
This includes photon exchange only inside the light cone, $vt>r,$ and vacuum fluctuations for all values of $t$ and $r$, being $r=x_B-x_A$ the distance between the qubits.  The remaining terms are
\begin{eqnarray}
A & \!\! = \!\! & \frac{1}{2}\bra{0}T(\mathcal{S}_A^+ \mathcal{S}_A^- +
\mathcal{S}_B^-\mathcal{S}_B^+)\ket{0}\label{e}\\
U_A & \!\! = \!\! & \bra{1}\mathcal{S}^-_A\ket{0},
V_B = \bra{1}\mathcal{S}^+_B\ket{0} , \nonumber\\
F & \!\! = \!\! & \frac{1}{2}\bra{2}T(\mathcal{S}_A^+ \mathcal{S}_A^-
+\mathcal{S}_B^-\mathcal{S}_B^+)\ket{0} \! , \, G = \bra{2}T(\mathcal{S}^+_B \mathcal{S}^-_A)\ket{0}.\nonumber
\end{eqnarray}
Here, $A$ describes intra-qubit radiative corrections, while $U_A, V_B, F$ and $G$ correspond to single-photon emission events by one or more qubits.

Note that virtual terms like $V$, $F$ and $G$, which do not conserve energy, are relevant only at very short times and are always neglected in the RWA. Here, we are interested in the short-time behavior, and therefore all the terms must be included \cite{power97,milonni95,biswas90}. Furthermore, only when the non-RWA terms are included, can it be said properly that the probability of excitation of qubit $B$ is completely independent of qubit $A$ when $r>vt,$ which is precisely the condition for microcausality discussed before \cite{milonni95,biswas90}.

The coefficients in Eq.~(\ref{wavefunction}) can be computed analytically~\cite{conjuan} as a function of two dimensionless parameters, $\xi$ and $K.$ The first one, $\xi=vt/r,$ was introduced before and it distinguishes the two different spacetime regions [Fig.~\ref{fig:setup}a], before and after photons can be exchanged. The second parameter is a dimensionless coupling strength
\begin{equation}
  K=\frac{4d^2N}{\hbar^2 v} = 2\left(\frac{g}{\Omega}\right)^2\label{g}.
\end{equation}
Note that the qubit-line coupling $g=d\sqrt{N\Omega}/\hbar$ corresponds to the qubit-cavity coupling that appears by taking the same transmission line and cutting to have a length $L=\lambda$ thus creating a resonator Refs.~\cite{wallraff04,blais04}. This formulation has the advantage of being valid both for inductive and capacitive coupling, the details being hidden in the actual expressions for $d$ and $N.$

Tracing over the states of the field, we arrive at the following
reduced density matrix
\begin{eqnarray}
\rho_{AB}=\frac{1}{c}\left( \begin{array}{c c c c}
\rho_{11}&0&0&\rho_{14} \\
0&\rho_{22}&\rho_{23}&0\\
0&\rho_{23}^*&\rho_{33}&0\\
\rho_{14}^*&0&0&\rho_{44}
\end{array}\right) , \label{state}
\end{eqnarray}
representing the two-qubit state in the basis formed by $\ket{ee},$
$\ket{eg},$ $\ket{ge},$ and $\ket{gg}.$ The coefficients with the
leading order of neglected contributions are
\begin{eqnarray}
\rho_{11}&=&|V|_B^2+\mathcal{O}(d^4),~
\rho_{22}=1+2\mathrm{Re}(A)+\mathcal{O}(d^4)\nonumber\\
\rho_{33}&=&|X|^2+|G|^2+\mathcal{O}(d^6),~
\rho_{44}=|U|_A^2+\mathcal{O}(d^4) \nonumber\\
\rho_{14}&=&U_A^*V_B+\mathcal{O}(d^4)=\langle0|\mathcal{S}_A^+ \mathcal{S}_B^+|0\rangle +\mathcal{O}(d^4)\label{j}\\
\rho_{23} &=& X^*+\mathcal{O}(d^4) , \nonumber
\end{eqnarray}
and the state is normalized $c=\sum_i \rho_{ii}.$

Let us now remark the validity of the perturbative methods applied in this work. The leading corrections to $C(\rho_{AB})$ (see Eq. (\ref{l}) below) come from the leading order corrections to $\rho_{23}, \rho_{11}, \rho_{44}$ (Eq.~(\ref{j})). In the case of $\rho_{23}$ we have $\rho_{23}(d^4)=(1+A) X^*+FG^*$ and $\rho_{23}(d^6)=\rho_{23}(d^4)+X_1+X_2$, where $X_1$ comes from the interference of one and two photon exchange amplitudes and $X_2$ comes from  the probability amplitude of three photon exchange. A rough upper bound for these two terms is given by $2\,|X|^3$.  For $\rho_{11}$ and $\rho_{44}$ they involve a number of photon emissions and reabsorptions by the same atom or by the other, giving a term $\rho_{11}\rho_{44} (d^6)=|U_A|^2|V_B|^2+A_1+A_2$, where rough upper bounds to $A_1$ and $A_2$ are  $2 |A||U_A|^2|V_B|^2$ and  $2|X||U_A|^2|V_B|^2 $ respectively . All these products are shown to be small for the regions of interest discussed here, $\xi < 2.$ The same techniques can be extended to all orders in perturbation theory since the bounds to the different contributions can be grouped and treated as power series, giving rise to corrections that remain negligible as long as $|A|$, $|X|$, $|U_A|^2$ and $|V_B|^2$ are small enough, like in the parameter range explored in this work. Finally, note that similar calculations and results can be obtained in the case in which the qubits have close but different frequencies.

\section{Entanglement dynamics and single photons}

We will use the concurrence $C$ to compute the entanglement of this state,
which is given by
\begin{equation}
C(\rho_{AB})=\frac{2}{c}\,\mbox{max}\left\{|\rho_{23}|-\sqrt{\rho_{11}\rho_{44}}\,,|\rho_{14}|-\sqrt{\rho_{22}\rho_{33}}\,,0\right\}\label{l}.
\end{equation}

Since all quantities depend only on two dimensionless numbers, $\xi$ and $K,$ we can perform a rather exhaustive study of the dynamics of entanglement between both qubits. To cover the widest possible spectrum of experiments, we have chosen coupling strengths over two orders of magnitude, $K/K_0= 1, 10, 100, 1000.$ The smallest value $K_0= 1.5\cdot10^{-4}$, which corresponds to $g/\pi\simeq175$ MHz and $\Omega/2\pi\simeq10$ GHz, that is for instance a charge qubit in the strong coupling limit with a transmission line~\cite{blais04}. The largest value, $K=1000\,K_0$ corresponds to $g\simeq2\pi\times 500$ MHz and $\Omega \simeq 2\pi\times 2$ GHz, and typically corresponds to a flux qubit directly coupled to a transmission line~\cite{bourassa09,peropadre09}, as shown in Fig.~\ref{fig:setup}b-c. Note that in this case, by building the qubit directly on the line, much larger couplings can be achieved, in the range of $1-4$ GHz, with $800$ MHz being recently obtained~\cite{forn}.
\begin{figure}
\includegraphics[width=0.85\linewidth]{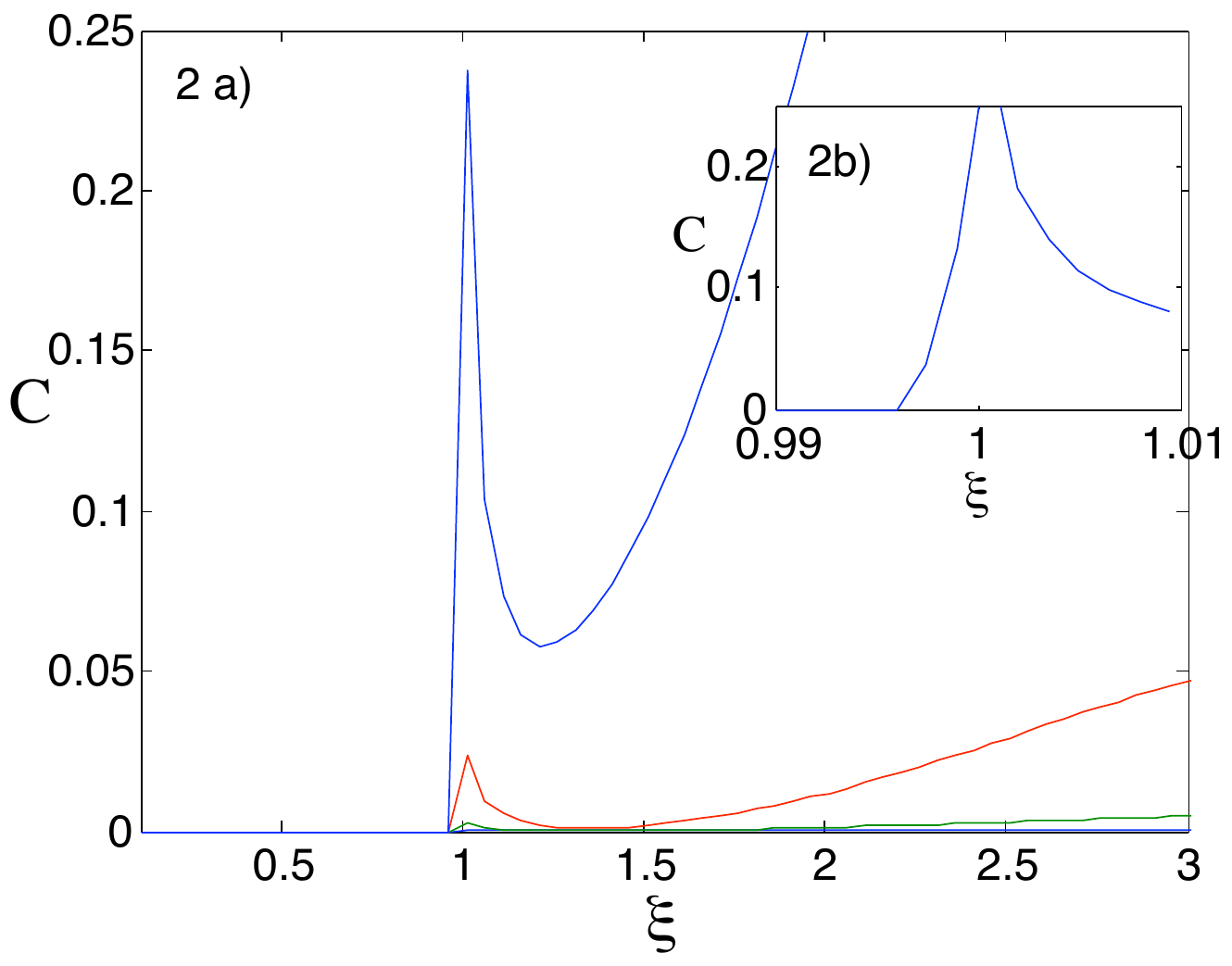}
\caption{ a) Concurrence vs. dimensionless separation $\xi$ for $r =
  \pi v / 4\Omega\sim \lambda/8$ and couplings $K= K_0$ , $10 K_0$ ,
  $100 K_0$ and $1000 K_0$ (bottom to top) b) Zoom around
  $\xi=1$ for the strongest coupling $K= 1000 K_0$. (Color online). }
\label{fig:concurrence}
\end{figure}

In Fig.~\ref{fig:concurrence} we plot the value of the concurrence for two qubits which are separated a distance $r=\lambda/8,$ using the couplings discussed before. Note how the entanglement jumps discontinuously to a measurable value right inside the light cone $(\xi > 1),$ signalling the arrival of photons. Furthermore, even a certain amount of entanglement appears outside the light cone, before photons could be exchanged. This is best seen for the largest couplings, as Fig.~\ref{fig:concurrence}b illustrates.

\begin{figure}[t]
\includegraphics[width=0.85\linewidth]{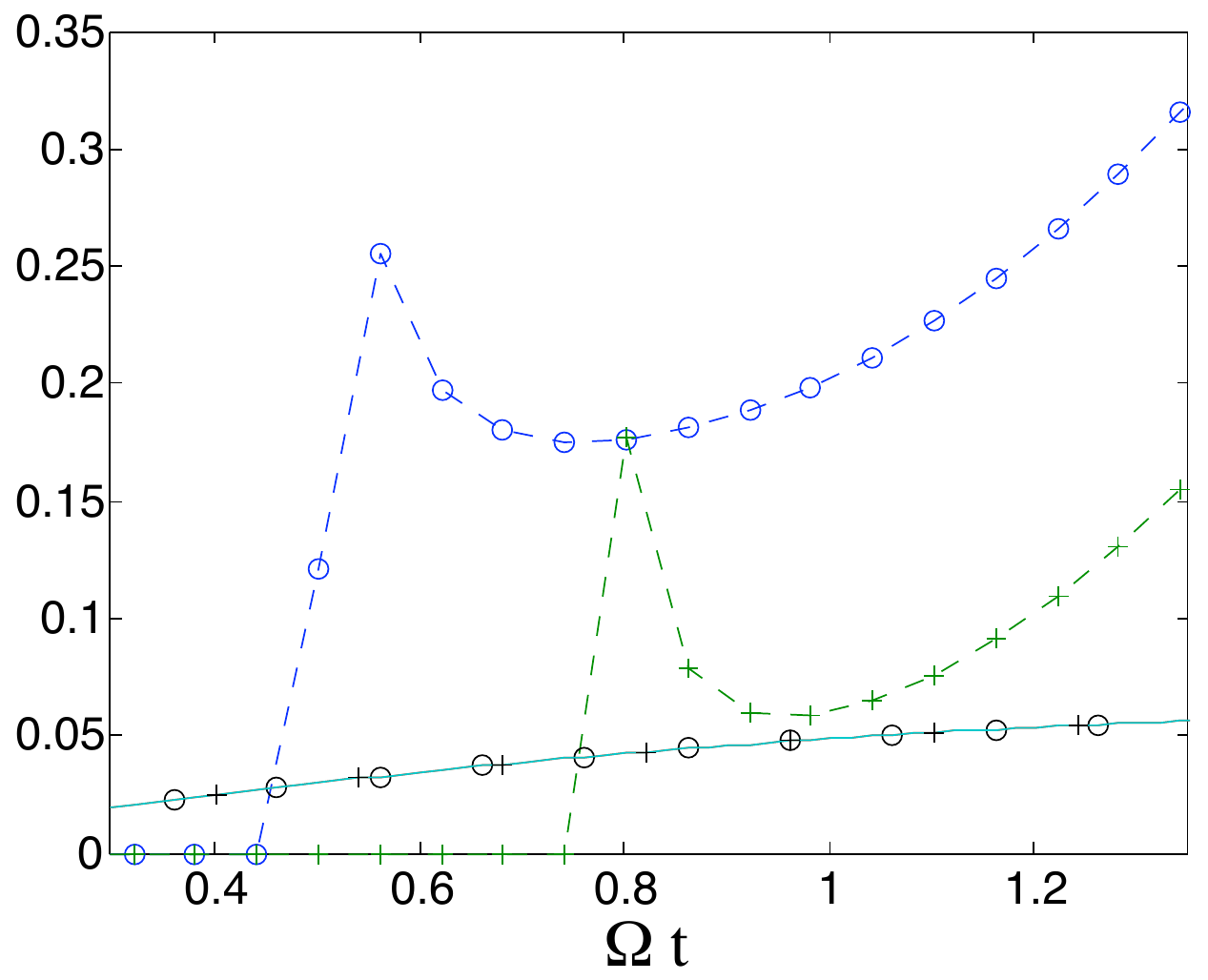}
\caption{Concurrence (dash) and probability of excitation of atom B (solid) vs. dimensionless time, $\Omega t.$ Qubits are separated by $r=\lambda/12$ (circles) and $\lambda/8$ (crosses) and have a coupling strength $K= 1000 K_0.$ Note that, following microcausality, the excitation probabilities do not depend on the separation $r$ outside the light cone (Color online).}
\label{fig:concurrence-exp}
\end{figure}

The dynamics looks even more exciting when we go back to lab time and space.  Fig.~\ref{fig:concurrence-exp} shows the concurrence and the excitation probability of qubit B, $p_B=|V_B|^2/c+ {\cal O}(d^4),$ for two different separations, $r=\lambda/12$ and $r=\lambda/8.$ The probability of excitation appears as independent of the qubit separation. This is exactly the case for the lowest order considered here, which only accounts for B self-interaction, and at all orders in perturbation theory ~\cite{power97} outside the light cone of this setup (region I in Fig.~\ref{fig:setup}a). This is in full agreement with microcausality. However, as can be seen in  Fig.~\ref{fig:concurrence-exp}, what was a tiny concurrence jumps to a sizable value when crossing the light cone $\Omega\, t = 2\pi/12$ and $\Omega\, t = 2\pi/8$. In other words, from the experimental point of view, it is the entanglement between the qubits and not the excitation probability $p_B$ what best signals the presence of a light cone and a finite propagation speed.

\section{Experimental implementation}

In order to study the dynamics of quantum correlations between the two superconducting qubits, one has to perform a partial or full tomography of their state. In the first and simpler case, performing measurements in different basis should be enough to gather an entanglement quantifier, such as a Bell inequality violation or, as studied in this paper, the concurrence. This has to be repeated many times, not only to gather sufficient statistics, but also to resolve different of instants of time before and after the light-cone boundary. This may seem a daunting task, but thanks to the speed at which quantum circuits operate and their fast repetition rate, it will be as demanding as recent experiments realizing a controlled-NOT gate~\cite{mooij07} or full two-qubit tomography~\cite{bialczak09}.

The actual experimental challenge, though, arises from the need to perform quantum measurements of the qubit state and ensuring that this state is not altered by the ongoing dynamics. One possibility is to perform very fast measurements of the qubits, which means faster than $1/\Omega.$ The typical response of measurement apparatus, which in the case of SQUIDs is around a few nanoseconds, sets an upper limit on the qubit and photon frequencies of a few hundreds of megahertzs, though we expect this to be improved in the near future.

Another more reliable approach is to connect and disconnect the coupling between the qubit and the transmission line. In this manner, we could prepare, entangle, and finally measure the qubits without interference or decay processes. If we work with flux qubits, a simple approach is to apply a very large magnetic flux on both qubits, taking the qubit away from its symmetry point. From a mathematical point of view, this amounts to adding a large contribution $E \sigma^x_{A,B}$ to the Hamiltonian. If done very quickly, the field projects the qubit on the same basis on which the coupling operates, eliminating the possibility of spontaneous emission. One would still need to combine the switching of this flux with short pulses that rotate the qubit basis in order to perform a complete set of measurements. The last and most elegant possibility is to effectively switch off all couplings between the qubit and the surrounding field. This can be achieved using a direct coupling between the qubit and the transmission line, with an scheme that incorporates an intermediate loop~[Fig.~\ref{fig:setup}c]. As we have shown in a recent work~\cite{peropadre09}, the result is a coupling that can be rotated and completely deactivated in a time of about 0.1 ns, that is the time needed to inject flux through the loop. The advantage is that, contrary to the case of a large external flux, the influence of the line is completely suppressed and makes it possible to easily rotate the qubits to perform all needed measurements.

\section{Conclusions}

Summing up, in this work we have proposed a circuit-QED experiment to study the dynamics of entanglement between two qubits that interact by exchanging traveling photons. Our work focuses on the existence of a finite propagation speed, the appearance of a light cone, the notion of microcausality and the possibility of achieving entanglement both by means of the correlated fluctuations of the vacuum and by photon exchange. The resulting predictions have a wide interest that goes beyond the assessment of microcausality in the QED of quantum circuits, demonstrating that the open transmission line is a useful mediator of entanglement, much like cavities and zero-dimensional resonators. Furthermore, the experiment we propose is also among the simplest ones that can probe the effective QFT for waveguides, both asserting the existence of propagating single photons and probing the dispersion relation at the single-photon level.  Finally, we have shown that entanglement via traveling photons works better for stronger qubit-line couplings, making it one of the first potential applications of the ultrastrong coupling regime \cite{bourassa09}.

\section*{ACKNOWLEDGEMENTS}

The authors thank P. Forn-D{\'\i}az for interesting discussions. E.S. acknowledges support from UPV-EHU GIU07/40, Ministerio de Ciencia e Innovaci\'on FIS2009-12773-C02-01, EuroSQIP and SOLID European projects. J.J.G.-R. thanks funding from Spanish MEC Project FIS2006-04885 and CSIC Project 200850I044. C.S and J.L acknowledge partial support from Spanish MEC Project FIS2008-05705.

\appendix*

\section{}
In this appendix we will give further details on the computations of the relevant magnitudes $|X|$, $|U_A|^2$, $|V_B|^2$ which are necessary to compute the concurrence Eq.(\ref{l}). With Eqs. (\ref{f})-(\ref{exchange}) and the commutation relations below Eq. (\ref{field}):
\begin{eqnarray} 
X=\frac{d^2\,N\,v}{\hbar^2}\int_{-\infty}^{\infty} dk |k|\,(e^{ikr}I_{t1}+e^{-ikr}I_{t2}) \label{exchange1}
\end{eqnarray}
with 
\begin{eqnarray}
I_{t1,2}=\int^{t}_0 dt_{2,1}\int^{t_{2,1}}_0dt_{1,2}e^{i\Omega\,(t_2-t_1)}e^{-iv|k|(t_{2,1}-t_{1,2})}.\label{exchange2}
\end{eqnarray}
Notice that the term with $I_{t2}$ gives the non-RWA probability amplitude associated to a single photon emission of qubit B followed by an absorption of qubit A. Performing the time integrations, inserting them in Eq. (\ref{exchange1}) and after some algebra, $X$ can be given as a combination of integrals of the form:
\begin{eqnarray}
\int_0^\infty\,d\,k\,\frac{\cos(\,k\,\gamma)}{k\,+\,\beta}&=&-\sin(\gamma\beta) si(\gamma\beta) -\cos(\gamma\beta) ci(\gamma\beta)\nonumber\\
\int_0^\infty\,d\,k\,\frac{\cos(\,k\,\gamma)}{k\,-\beta}&=&-\sin(\gamma\beta) si(\gamma\beta) -\cos(\gamma\beta) ci(\gamma\beta)\nonumber\\&-&\pi\sin(\gamma\beta)\label{integrals}\\
\int_0^\infty\,d\,k\,\frac{\sin(\,k\,\gamma)}{k\,+\,\beta}&=&\sin(\gamma\beta) Ci(\gamma\beta) -\cos(\gamma\beta) si(\gamma\beta)\nonumber\\
\int_0^\infty\,d\,k\,\frac{\sin(\,k\,\gamma)}{k\,-\,\beta}&=&-\sin(\gamma\beta) Ci(\gamma\beta) +\cos(\gamma\beta) si(\gamma\beta)\nonumber\\&+&\pi\cos(\gamma\beta)\nonumber
\end{eqnarray}
with $\gamma,\beta>0$ and the conventions in \cite{bateman} for the $si$ and $Ci$. Putting all together we find:
\begin{eqnarray}
&&X=\frac{K}{2}(i\pi  \rho\xi \sin(\rho)-e^{ \frac{i\rho\xi}{2}} ((1+i \tau_{-})(C(\tau_{-})-S(\tau_{-})\nonumber\\&&-\pi\sin(\tau_{-})\Theta(1-\xi) )+(1-i\tau_{+})(-C(\tau_{+})-S(\tau_{+})\nonumber\\&&-\pi\sin(\tau_{+}))+)(-\tau_{-}+i)(-SC(\tau_{-})+CS(\tau_{-})+\pi\nonumber\\&&\cos(\tau_{-}))+(-\tau_{+}-i)(-SC(\tau_{+})+CS(\tau_{+})+\pi \cos(\tau_{+}))\nonumber\\&&-2)-e^{\frac{-i \rho\xi}{2}} ((1-i \tau_{-})(-C(\tau_{-})-S(\tau_{-})-\pi\sin(\tau_{-})\nonumber\\&&\Theta(\xi-1))+(1+i\tau_{+})(-C(\tau_{+})-S(\tau_{+}))+(\tau_{-}+i)\nonumber\\&&(SC(\tau_{-})-CS(\tau_{-}))+(\tau_{-}-i)(SC(\tau_{+})-CS(\tau_{+})+\pi\nonumber\\&&\cos(\tau_{+}))-2) - 2 - 2C(\rho)-2 S(\rho)-\rho(-2SC(\rho) \nonumber\\
&&+2 CS(\rho))\label{exchange3}
\end{eqnarray}
where $\xi$ has been defined in the main text, $\rho=\Omega\,r/v$ is a dimensionless distance, $\tau_{-}=\rho(1-\xi)=\rho-\Omega\,t$, $\tau_{+}=\rho (1+\xi)=\rho+\Omega\,t$, and we define $C(x)=\cos(x)\,Ci(x)$, $S(x)=\sin(x)\,si(x)$, $CS(x)=\cos(x)\,si(x)$, $SC(x)=\sin (x)\,Ci(x)$. Notice the dependence with the spacetime region through the factors with the Heaviside function $\Theta$.

Now we come to the emission probabilities $|U_A|^2$, $|V_B|^2$, which are given by
\begin{equation}
|U_A|^2  \!\! =  \bra{0}\mathcal{S}^+_A\mathcal{S}^-_A\ket{0},
|V_B|^2 = \bra{0}\mathcal{S}^-_B\mathcal{S}^+_B\ket{0}
\end{equation}
Following similar techniques we find that $|U_A|^2=f_{+}(\Omega\,t)$, $|V_B|^2=f_{-}(\Omega\,t)$, where:
\begin{equation}
f_{\pm}=\frac{K}{2} (\pi\,\Omega\,t\pm2 (\cos(\Omega\,t)+ \Omega\,t \,Si(\Omega\,t)-1))
\end{equation}
where $Si$ must not be mistaken by $si$, $Si=si\,+\pi/2$ as usual.

Finally, notice that $A$ (Eq.(\ref{e})) is a sum of two terms like $|U_A|^2$ and $|V_B|^2$ with the time ordering operator $T$ and that $U_A*V_B$ (Eq.(\ref{j})) is similar to $X$ without $T$.

\end{document}